%% file: ms.tex
\documentclass{emulateapj}

\usepackage{comment}
\usepackage{lscape}
\usepackage{rotating}

\slugcomment{Astrophysical Journal 712 (2010) 942-950}

\shorttitle{Dust in $z>4$ submillimeter galaxies}
\shortauthors{Micha{\l}owski et al.}

\newcommand{\sfrircol}{5}

\newcommand{\mburstcol}{9}

\newcommand{\mdustcol}{11}

\newcommand{\avcol}{14}

\newcommand{\submm}{submillimeter}

\begin{document}

\title{Rapid dust production in submillimeter galaxies at $z>4$?
}

\author{Micha{\l}~J.~Micha{\l}owski\altaffilmark{1,2}
, 
Darach Watson\altaffilmark{1},
and 
Jens Hjorth\altaffilmark{1}
	}

\altaffiltext{1}{Dark Cosmology Centre, Niels Bohr Institute, University of Copenhagen, Juliane Maries Vej 30, DK-2100 Copenhagen \O, Denmark; {\tt michal@dark-cosmology.dk}}
\altaffiltext{2}{%
{Scottish Universities Physics Alliance}, Institute for Astronomy, University of Edinburgh, Royal Observatory, Edinburgh, EH9 3HJ, UK}

\begin{abstract}

The existence of {\submm}-selected galaxies (SMGs) at redshifts $z>4$ has recently been confirmed. Using simultaneously all the available data from UV to radio we have modeled the spectral energy distributions of the six known spectroscopically confirmed SMGs at $z>4$. 
We find that their star formation rates (average $\sim2500\,M_\odot$ yr$^{-1}$), stellar ($\sim3.6\times10^{11}\,M_\odot$) and dust ($\sim6.7\times10^{8}\,M_\odot$) masses, extinction ($A_V\sim2.2$ mag), and gas-to-dust ratios ($\sim60$) are within the ranges for $1.7<z<3.6$ SMGs. 
Our analysis suggests that  
infrared-to-radio luminosity ratios of SMGs do not change up to redshift $\sim5$ and are lower by a factor of $\sim2.1$ than the value corresponding to the local IR-radio correlation.
However,  we also find dissimilarities between $z>4$ and lower-redshift SMGs. Those at $z>4$ tend to be among the most star-forming, least massive and  hottest ($\sim60$ K) SMGs and exhibit the highest fraction of stellar mass formed in the ongoing starburst ($\sim45$\%). This indicates that at $z>4$ we see earlier stages of evolution of {\submm}-bright galaxies. 
Using the derived properties for $z>4$ SMGs we investigate the origin of dust at epochs less than $1.5$ Gyr after the big bang. 
This is significant to our understanding of the evolution of the early universe.
For three $z>4$ SMGs, asymptotic giant branch stars could be the dominant dust producers. However, for { the remaining three} 
only supernovae (SNe) are  efficient and fast 
enough to be responsible for dust production, 
though requiring a very high dust yield per SN ($0.15$--$0.65\,M_\odot$).
The required dust yields are lower if
a top-heavy initial mass function or significant dust growth in the interstellar medium 
is assumed. 
We estimate lower limits of the contribution of SMGs to the cosmic star formation and stellar mass densities at $z\sim4$--$5$ to be $\sim4$\% and $\sim1$\%, respectively. 
\end{abstract}


\section{Introduction}
\label{sec:intro}

Submillimeter-selected galaxies (SMGs) are among the most powerful starburst galaxies in the Universe. Most of them have been found at redshifts $1.5$--$3$ \citep{chapman05}. Their complex selection criteria \citep{blainT}, in particular the requirement of a radio detection to obtain a precise localization, make it difficult to discover the very high redshift tail of SMGs.
{This was addressed using  deep, high resolution observations of SMGs 
 \citep{iono06,tacconi06,tacconi08,wang07,wang09,younger07,younger08,younger08b,younger09,younger09b,dannerbauer08,cowie09}.}
 The existence of SMGs at $z>4$ has recently been { spectroscopically} confirmed by \citet{coppin09}, \citet{capak08}, \citet{schinnerer08}, \citet{daddi09,daddi09b}, and \citet{knudsen08b,knudsen09}.

At these redshifts the age of the Universe is $<1.5$~Gyr, which enforces the need for careful analysis of the timescales for  formation of stars and dust. 
The important question is if supernovae (SNe), or asymptotic giant branch (AGB) stars, or some other sources are responsible for production of dust residing in these galaxies. Locally, dust is predominantly formed by evolved, post-main-sequence stars \citep{gehrz89}, but the situation may be different at high redshifts. \citet{dwek07} claimed that only SNe can produce dust on timescales less than $1$ Gyr, but it has been shown by \citet{valiante09} that 
 AGB stars begin to dominate dust production over SNe as early as $150$--$500$ Myr after the onset of star formation \citep[see also][]{sloan09}. SN-origin dust has been claimed to be present in $z\sim6.2$ quasar \citep{maiolino04} and in two gamma-ray burst host galaxies at $z\sim6.3$ (\citeauthor{stratta07b} \citeyear{stratta07b}, but see other interpretation of their data in \citeauthor{zafar10}  \citeyear{zafar10}) and at $z\sim5$ \citep{perley09}.
 
 In order to understand the formation of SMGs and their evolution  through cosmic time, it is also important to compare high- and low-redshift  SMG samples. This may help to constrain when their stars were formed.

SMGs at $z>4$ are also suitable to study the infrared (IR) - radio correlation. This remarkably tight correlation, found locally \citep{helou85,condon}, was
{ studied at redshifts $z\lesssim3.5$ } \citep{garrett02,gruppioni03,appleton04,kovacs06,boyle07,marleau07,vlahakis07,yang07,beswick08,hainline08phd,ibar08,sajina08,garn09,michalowski10smg,murphy09, murphy09b,rieke09,seymour09,younger09,sargent10}.
No significant evolution of the correlation was found up to these redshifts, { but SMGs seem to form a correlation by their own offset toward higher radio luminosities \citep{kovacs06,michalowski10smg,murphy09,murphy09b}.
The only sign of evolution was reported by \citet{ivison09} based on stacking analysis of $24\,\micron$-selected galaxies, though possibly interpreted as a selection effect.}
 It is however possible that the correlation breaks down at even earlier epochs due to 
changes in star formation processes, e.g.,
suppression of radio emission in inverse Compton losses off the CMB photons as suggested by \citet{lacki09}, \citet{lacki09b}, and \citet{murphy09b}.

In \citet{michalowski10smg} we analyzed the full UV-to-radio spectral energy distributions (SEDs) of 76 SMGs from the \citet{chapman05} sample with spectroscopic redshifts up  to $z<3.6$. Here we extend that study by analyzing the sample of all spectroscopically confirmed SMGs at $z>4$.
The main objective of this paper is to characterize the required efficiency of dust producers (SNe and AGB stars) at these early epochs of the evolution of the Universe.
 In Section~\ref{sec:sample} the SMG sample is presented. We outline our methodology and derive the properties of SMGs in Section~\ref{sec:results} and discuss the implications in Section~\ref{sec:discussion}. Section~\ref{sec:conclusion} closes with our conclusions.
We use a cosmological model with $H_0=70$ km s$^{-1}$ Mpc$^{-1}$,  $\Omega_\Lambda=0.7$ and $\Omega_m=0.3$.

\input{phottab}
\input{limtab}

\section{Sample}
\label{sec:sample}

\begin{figure*}
\begin{center}
\includegraphics[width=\textwidth, viewport=28 200 560 820, clip]{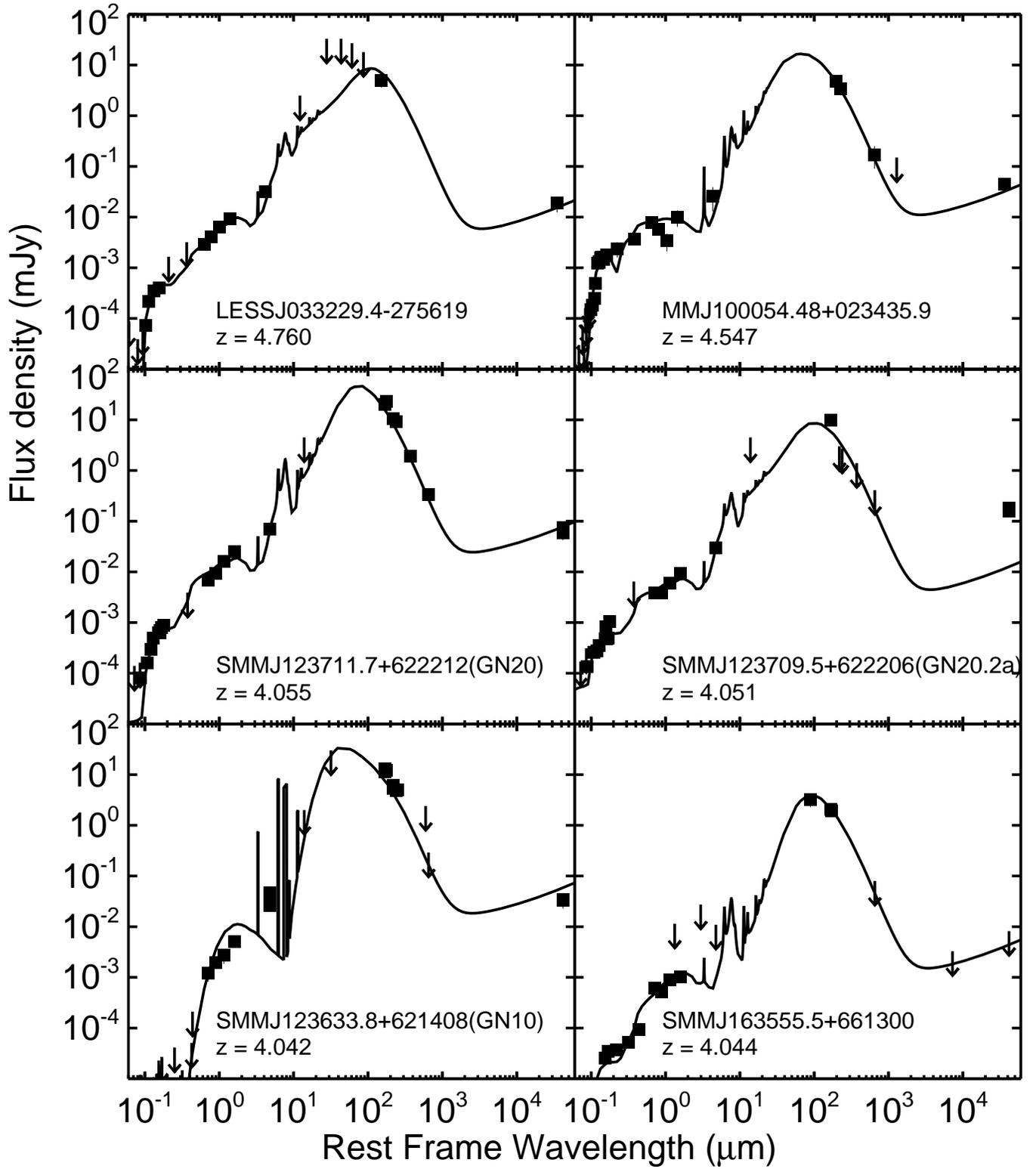}
\end{center}
\caption{Spectral energy distributions (SEDs) of $z>4$ SMGs. {\it Solid lines}: the best GRASIL fits. 
{\it Squares}: detections with errors, in most cases, smaller than the size of the symbols. {\it Arrows}: $3\sigma$ upper limits (values marked at the base). The data for \protect\object{SMMJ163555.5+661300} have been corrected for lensing magnification of a factor of $5.5$.
}
\label{fig:sed}
\end{figure*}

We selected all six SMGs with robust (optical or CO) redshifts at $z>4$ identified by \citet{coppin09}, \citet{capak08}, \citet{schinnerer08}, \citet{daddi09,daddi09b}, and \citet{knudsen08b,knudsen09} in the ECDF-S \citep[$900$ arcmin$^2$;][]{coppin09,greve09,weiss09b}, COSMOS \citep[$1080$ arcmin$^2$;][]{scott08}, GOODS-N \citep[$100$ arcmin$^2$;][]{hughes98,barger00,chapman01,borys03,serjeant03,wang04}, and Abell 2218 \citep[$11.8$ arcmin$^2$;][]{knudsen06} fields. The photometric data are presented in Tables~\ref{tab:phot} and \ref{tab:lim}. The data for \object{SMMJ163555.5+661300} have been corrected for lensing magnification of a factor of $5.5$ obtained by \citet{knudsen09} using the model of \citet{eliasdottir07}.

Our sample is not homogeneously selected. Namely, some of the sources are bright enough in the optical to allow spectroscopy, whereas redshifts of some of them were measured based on CO emission (one even not detected at optical wavelengths). In particular, \object{LESSJ033229.4-275619} and \object{MMJ100054.48+023435.9} were targeted spectroscopically, because they are $V$-dropouts suggesting $z\approx5$ \citep{coppin09,capak08} and  their optical counterparts (and hence, redshifts) are based on detections at radio wavelengths, which likely biases them toward high star formation rates (SFRs). The CO lines from \object{SMMJ123711.7+622212} \citep[called GN20 in][]{pope06} and \object{SMMJ123709.5+622206} (GN20.2a)  were detected serendipitously while observing an angularly close galaxy at $z=1.522$ \citep{daddi09}, and the CO emission of \object{SMMJ123633.8+621408} (GN10) was searched for under the assumption that it is a member of the protocluster structure containing GN20 and GN20.2a \citep{daddi09b}. Finally, \object{SMMJ163555.5+661300} was detected because of its lensing magnification and is therefore intrinsically the faintest member of our sample.

It is therefore not easy to quantify how selection effects influence our results. Very likely our sample is biased toward high luminosity objects, i.e., with high SFRs. This is supported by the fact that  \object{SMMJ163555.5+661300}, magnified by lensing { and the only member of our sample not detected in the radio}, has a much lower SFR than blank-field members of our sample (Table~\ref{tab:grasilres}). However, given the significance of the {\it Spitzer} IRAC detections (rest-frame $\sim1$--$2\,\micron$) of $\gtrsim10\sigma$, our sample is {\em not} biased against low stellar masses.

\section{SED Fitting and Results}
\label{sec:results}

\input{grasilrestab}

We applied the SED fitting method detailed in \citet[][see therein a discussion of the derivation of galaxy properties and typical uncertainties]{michalowski08,michalowski09,michalowski10smg} based on 35\,000 templates in the library of \citet{iglesias07}, plus some templates of \citet{silva98} and \citet{michalowski08}, all developed in GRASIL \citep{silva98}.
The templates cover a broad range of galaxy properties and were tested to reproduce the SEDs of high-redshift galaxies \citep{silva98,iglesias07,michalowski08,michalowski10smg}. { Their star formation histories are assumed to be a smooth Schmidt-type law \citep[SFR proportional to the gas mass to some power, see][for details]{silva98} with a starburst (if any) on top of that starting $50$ Myr before the time at which the SED is computed. There are seven free parameters in the library of \citet{iglesias07}: the normalization of the Schmidt-type law, the timescale of the mass infall, the intensity of the starburst, the timescale of the molecular cloud destruction, the optical depth of molecular clouds, the age of a galaxy and the inclination of a disk with respect to the observer.

We scaled all the SEDs to match the data (detections) and chose the one with the lowest $\chi^2$. In all but one case (see below) the upper limits did not provide a constraint on the best-fitting model.
}

In all but one case we obtained the best fits using the library of \citet{iglesias07}. For \object{GN10} we fitted a model corresponding to a $0.1$ Gyr old progenitor of an elliptical galaxy \citep{silva98} with modified maximum grain temperatures from $400$ K to $100$ K in order to suppress strong mid-IR emission in the original model, otherwise inconsistent with the data. All other models failed to reproduce its extremely red observed-frame $2.1$--$3.6\,\micron$ color \citep{wang09}.

The best fits\footnote{The SED fits can be downloaded from\\ \protect\url{http://archive.dark-cosmology.dk}}  are shown in Figure~\ref{fig:sed}. It is apparent that GRASIL models can reproduce the SEDs of even such distant galaxies.
The resulting properties of the galaxies are listed in Table~\ref{tab:grasilres}.

{
In order to assess the uncertainties of the derived parameters, we calculated the confidence intervals using SED models with $\Delta\chi^2<1$. The resulting uncertainties are a factor of $\sim2$--$4$ for SFRs and $L_{\rm IR}$;  a factor of $\sim1.5$--$2.5$ for $M_*$;  a factor of $\sim2$--$5$ for $M_{\rm burst}$; $\sim10$ K for $T_d$; and $0.3$--$1.0$ mag for $A_V$. 
The uncertainty  on $T_d$ introduces a factor of $\sim1.5$--$2.5$ error on $M_d$ and allowing the emissivity $\beta$ to vary in the range $1$--$2$ results in an uncertainty of a factor of $<3$ on $M_d$.
The choice of a \citet{salpeter} initial mass function (IMF) with cutoffs of $0.15$ and $120\,{\rm M}_\odot$ introduces a maximum systematic error of a factor of $\sim2$ in the determination of the stellar masses and SFRs \citep{erb06}.
}

\begin{table*}
\caption{Properties of $z>4$ SMGs Derived by Other Authors \label{tab:others}}
\centering
\begin{tabular}{l cccc l}
\hline\hline
	& SFR & $L_{\rm IR}$ & $M_*$ & $M_d$\\
SMG & ($M_\odot$ yr$^{-1}$) & ($10^{13}L_\odot$) & ($10^{11}M_\odot$) & ($10^{9}M_\odot$) & Ref.\\
\hline
LESS & 1000 & 0.61 & 0.5 & 0.5 &  \citet{coppin09} \\ 
MM & 1000-4000 & 0.5--2.0  & 0.1 & 	& \citet{capak08}, \citet{schinnerer08}\\
GN20 &		& 1.0--5.0 & 2.3 & 	& \citet{daddi09}, \citet{younger08b}, \citet{casey09}\\
GN20.2a & 	& 0.5--1.6		&0.5 & & \citet{daddi09}, \citet{casey09}\\
GN10 & 2400 & 1.2--2.5 & 1.0--3.0 & & \citet{daddi09b}, \citet{wang09}\\
SMM & 230 & 0.13 & 0.16 & & \citet{knudsen09} \\
\hline
\end{tabular}
\tablecomments{The first column lists all SMGs in the order given in Table~\ref{tab:grasilres}. Only the first parts of their names are given for brevity.}
\end{table*}

Our estimates of SFRs, $L_{\rm IR}$ and  $M_d$ are consistent within a factor of $<3$ with those obtained by \citet{coppin09}, \citet{capak08}, \citet{schinnerer08}, \citet{younger08b}, \citet{casey09}, \citet{daddi09,daddi09b}, \citet{knudsen09}, and \citet{wang09} {(compare Tables~\ref{tab:grasilres} and \ref{tab:others}}). Our $M_*$ estimates are consistent with those of  \citet{daddi09,daddi09b}, and \citet{wang09}  after taking into account that \citet{daddi09,daddi09b} used a \citet{chabrier03} IMF resulting in  stellar masses $1.8$ times lower than for the \citet{salpeter} IMF \citep{erb06}. 

However we obtained  stellar masses for \object{LESSJ033229.4-275619} and \object{MMJ100054.48+023435.9} $\sim10$ times larger than \citet{coppin09}%
\footnote{Note that our estimate agrees with that of \citet{stark07}. }
 and \citet{capak08}, respectively, and for \object{SMMJ163555.5+661300} $\sim5$ times larger than \citet{knudsen09}. The difference can be explained by the fact that \citet{coppin09} assumed a mass-to-light ratio of $M_*/L_K=0.1$ and the determinations of \citet{capak08} and \citet{knudsen09} correspond to $M_*/L_K\sim0.03$ and $\sim0.17$, respectively. These are very low values { compared to lower-redshift galaxies} \citep[e.g.,][]{drory04,portinari04,labbe05,castroceron06,castroceron09,vandervel06,courty07,michalowski10smg,savaglio09} giving  lower limits on stellar masses. On the other hand, we do not assume mass-to-light ratios, but derive them from the stellar population models incorporated in GRASIL.

\section{Discussion}
\label{sec:discussion}

\subsection{Formation of Stars in $z>4$ SMGs}
\label{sec:stars}

All $z>4$ SMGs in our sample are characterized by an extremely strong starburst episode (average SFR  $\sim2500\,M_\odot$ yr$^{-1}$, Column~{\sfrircol} of Table~\ref{tab:grasilres}) during which a substantial fraction  (average $\sim45$\%, Column~\mburstcol) of their stellar population was formed. They are therefore manifestations of the strongest known star-forming events in the universe.
Their high stellar masses (average $\sim3.6\times10^{11}\,M_\odot$) agree with a suggestion of \citet{dave09} based on numerical simulations that the most rapidly star forming galaxies coincide with the most massive galaxies.

The properties of the galaxies are within the ranges found by \citet{michalowski10smg} for the $1.7<z<3.6$ SMG sample \citep{chapman05}. However, $z>4$ SMGs tend to be among the most star-forming, least massive, and hottest SMGs and exhibit the highest fraction of stellar mass formed in the ongoing starburst. Namely, $43$\% of $1.7<z<3.6$ SMGs have lower SFRs than any of the $z>4$ SMGs%
\footnote{We exclude \protect\object{SMMJ163555.5+661300} from this analysis, because its unlensed {\submm} flux of $\sim2$ mJy makes it impossible to be detected by SCUBA in the blank-field survey similar to those used by \citet{chapman05}. }%
, whereas only $2$\% have higher SFRs; $30$\% of $1.7<z<3.6$ SMGs have higher stellar masses, whereas only $8$\% have lower stellar masses; $55$\% have { a} lower fraction of stellar mass formed in the ongoing starburst episode, whereas only $2$\% have higher fraction; and $28$\% have lower dust temperatures, whereas only $4$\% have higher temperature. This can be interpreted as SMGs at $z>4$ representing earlier stages of the formation of {\submm}-bright galaxies at which the pre-existing stellar population is less pronounced and therefore the ongoing starburst episode contributes more to the final stellar mass. 

However, we note that we cannot exclude the possibility that these galaxies are  AGN dominated,
{ suggested by their high { IR} luminosities and dust temperatures.}
In that case the SFRs and $M_*$ we derive would be upper limits.

If \object{GN10} is not AGN dominated, then the fact that we could only fit its SED using a template corresponding to a young ($0.1$ Gyr old) progenitor of an elliptical galaxy supports the hypothesis that SMGs evolve into ellipticals. According to our model,  after $1.5$ Gyr \object{GN10} will evolve into a massive elliptical containing $\sim10^{12}\,M_\odot$ stars. In order to explain its extremely red observed-frame $2.1$--$3.6\,\micron$ color, we did not need to invoke an old stellar population as suggested by \citet{wang09}. Its spectrum is reproduced by a  young stellar population residing in molecular clouds heavily obscured by dust with $A_V\sim7.8$ mag (Table~\ref{tab:grasilres}, Column~\avcol). This estimate is similar to the values obtained using only near-IR and optical data  by \citet{daddi09b} and \citet{wang09}.

On the other hand the remaining $z>4$ SMGs are only moderately obscured ($A_V\sim2$ mag; Column~{\avcol}). { We checked that this value of extinction is roughly consistent with the expectation based on the estimation of the  amount of energy reprocessed by dust. A crude estimate of the UV ($0.2\,\mu$m) extinction can be calculated using $A_{0.2\,\mu{\rm m}}=2.5\log(\mbox{SFR}_{\rm IR}/\mbox{SFR}_{\rm UV})$. This gives $A_{0.2\,\mu{\rm m}}\sim20$ mag for \object{GN10} and $\sim3.8$--$6.3$ mag for { the} other objects. Assuming that $A_{0.2\,\mu{\rm m}}\sim3\times A_V$ \citep[e.g.,][]{gordon03}, this corresponds to $A_V\sim6.6$ mag for \object{GN10} and $A_V\sim1.2$--$2.1$ mag for { the} other objects. These values are consistent with those derived from the SED modeling (Table~\ref{tab:grasilres}).
}

We estimated the SMG comoving volume densities 
{ as a sum of SFRs (or $L_{\rm IR}$, or stellar or dust masses) of $z>4$ SMGs%
\footnote{We again exclude \protect\object{SMMJ163555.5+661300} because it is lensed.}
 divided by the comoving volume in the redshift interval $4$--$5$ ($5.9\times10^6$ Mpc$^3$).
We obtained }
 $\rho_{\rm SFR}=2.6\times10^{-3}\,M_\odot \mbox{ yr}^{-1} \mbox{ Mpc}^{-3}$ \citep[a similar value was found by][based on candidates for high-redshift SMGs]{younger09b}, $\log\rho_{\rm LIR}=7.17 \,L_\odot \mbox{ Mpc}^{-3}$, $\log\rho_*=5.54\,M_\odot \mbox{ Mpc}^{-3}$, and $\log\rho_{\rm dust}=2.81\,M_\odot \mbox{ Mpc}^{-3}$. However one must keep in mind that these numbers could be affected by cosmic variance, because three out of five SMGs considered here are members of a protocluster structure \citep{daddi09,daddi09b}. The contributions to the cosmic SFR and stellar mass densities of SMGs at these redshifts are $4.4$\% and $1.0$\%, respectively, using the compilation of the total values in \citet[][Tables~A.4 and A.5]{michalowski10smg}. These numbers indicate that currently detected SMGs did not contribute significantly to the cosmic star formation history at $z>4$, but our estimates should be regarded as lower limits since more of such distant SMGs could still be undetected in the fields discussed here. 

\subsection{Producers of Dust in $z>4$ SMGs}

\input{dustyieldstab}

The dust masses we find for $z>4$ SMGs (Table~\ref{tab:grasilres}, Column~\mdustcol) are similar to those derived for $z\sim4$--$6$ quasars  \citep[a few$\times10^8 M_\odot$;][]{dunlop94,benford99,archibald01,omont01,priddey01,priddey03,priddey08,isaak02,bertoldi03,robson04,beelen06,wang08b,martinezsansigre09}. These huge dust masses  indicate that  dust in SMGs was efficiently formed and able to survive even when the age of the universe was only $1.2$--$1.5$ Gyr.

As detailed below we find that {\it i}) AGB stars are efficient and fast enough to form dust in { the} \object{LESSJ033229.4-275619}, \object{GN20} and \object{SMMJ163555.5+661300}; %
{\it ii})  only SNe are efficient and fast enough to form dust in \object{MMJ100054.48+023435.9}, {  \object{GN20.2a}}, and \object{GN10}, as long as the unusually high dust yields derived for  \object[Cassiopeia A]{Cassiopeia~A} and Kepler SN remnants are correct and typical, or if a top-heavy IMF and/or significant dust growth in the interstellar medium (ISM) is assumed. If these assumptions are correct, then SNe could also produce dust in SMGs mentioned in~{\it i}.

{ These conclusions are not significantly affected by the uncertainties reported in Section~\ref{sec:results}. Even if dust masses were higher or stellar masses lower by a factor of a few, the required dust yields per SN would not exceed that of  \object[Cassiopeia A]{Cassiopeia~A}  \citep{dunne09casA}  and Kepler \citep{gomez09}. It would however be more critical for AGB stars if dust masses were higher than we derived, since the required dust yields for  \object{LESSJ033229.4-275619}, \object{GN20}, and \object{SMMJ163555.5+661300} are already on the high end of the theoretically allowed yields. However, a decrease of dust masses by a factor of a few would not change our conclusion that AGB stars are not effective enough to form dust in \object{MMJ100054.48+023435.9},  \object{GN20.2a}, and \object{GN10}.
}

\subsubsection{Asymptotic giant branch stars}

In order to investigate whether AGB stars can be responsible for dust production in $z>4$ SMGs, we estimated (see the Appendix) the  average dust yields required per star with mass $2.5<M<8\,M_\odot$ and main-sequence lifetime in a range $1$ Gyr--$55$ Myr \citep[calculated as $10^{10}\,\mbox{yr} \times [M/M_\odot \mbox{$]$} ^{-2.5}$; e.g.,][]{kippenhahn90}. The lower mass limit was chosen to ensure that the stars considered can start producing dust within the age of the universe at the redshifts of our sources ($1.2$--$1.5$ Gyr). 

The results are listed in the first row of Table~\ref{tab:dustyields}. They are independent of the assumed star formation history of galaxies, but depend only on derived dust and stellar masses, assumed IMF and measured redshifts.
We find that each AGB star would need to produce $\sim0.03$--$0.07\,M_\odot$ of dust in order to explain the dust in $z>4$ SMGs, excluding \object{GN20.2a} (see below).
 These numbers are close to the highest theoretical dust yields of AGB stars \citep[][{ the total dust yields per AGB star are approximately metallicity independent}]{morgan03,ferrarotti06} making them plausible dust producers. Assuming a top-heavy IMF does not change this result significantly { (the required dust yields increase by $\sim30$\%), because the choice of the slope of the IMF affects the number of high- and low-mass stars, leaving the number of intermediate-mass stars approximately constant}. { For \object{GN20.2a} the required dust yield ($0.17\,M_\odot$) is too high to claim that AGB stars formed its dust.}

However, \object{MMJ100054.48+023435.9} and \object{GN10} formed the majority of their stars in the ongoing starburst episode (Column~{\mburstcol}, Table~\ref{tab:grasilres}), which is too short for the $2.5<M<8\,M_\odot$ stars considered above to finish their main-sequence phase. Therefore AGB stars could not contribute to the dust production in these two galaxies. To quantify this, we calculated the required dust yields for AGB stars taking into account only stars that were born before the ongoing starburst (replacing $M_*$ by $M_*-M_{\rm burst}$ in Equation \ref{eq:nstars}). The resulting yields (the second row of Table~\ref{tab:dustyields}) for \object{LESSJ033229.4-275619}, \object{GN20}, \object{GN20.2a}, and \object{SMMJ163555.5+661300} do not differ significantly from our previous estimate (the first row of Table~\ref{tab:dustyields}), because in this way we removed $\lesssim30$\% of stars (those formed during the ongoing starburst). However the yields for \object{MMJ100054.48+023435.9} and \object{GN10} become too high to claim that AGB stars formed dust in these galaxies.

A potential limitation of this claim is the uncertainty in determining the fraction of stellar mass formed during the ongoing starburst episode. If more stars were formed in the past, then AGB stars could be responsible for dust production in these two galaxies. 
{ However, even if it was the case, then the current SFRs would be the same as we derive, because they are fixed by strong {\submm} emission. Then the ongoing starburst becomes unfeasibly  short ($<25$ Myr) in order not to produce more stars than is inferred from the optical to near-infared part of the spectra.
}
Therefore, more stars in these two SMGs could have been formed before the ongoing starburst only if  the current SFRs are overestimated due to a significant AGN contribution.

\begin{figure*}
\begin{center}
\plotone{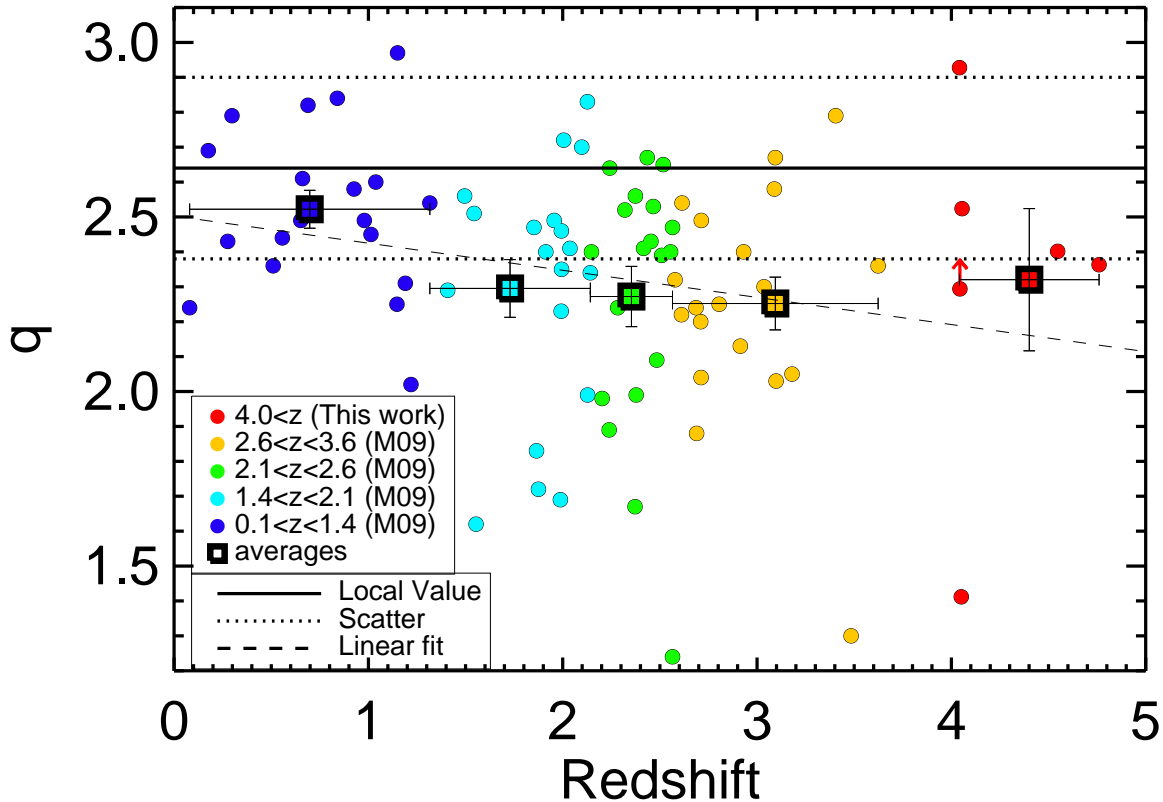}
\end{center}
\caption{Ratio of the infrared ($8$-$1000\,\micron$) and radio luminosities $q=\log(L_{\rm IR}/3.75\times10^{12}/I_{1.4})$ as a function of redshift of SMGs. Both SMGs at $z>4$ ({\it red circles}) and at lower redshifts ({\it yellow, green, blue, violet circles}, divided into four redshift bins) from \citet[][M09]{michalowski10smg} are shown.  
{ In the redshift range $1.4<z<5.0$ no significant evolution of the IR-radio correlation is found for SMGs. The average $q$ for SMGs ({\it squares}) is  however offset toward higher radio luminosities (factor of $\sim2.1$--$2.3$) from the local relation \citep[$q=2.64$: {\it solid line} with scatter, $0.26$: {\it dotted lines};][]{bell03}. }
{ A linear fit to the data ({\it dashed line}) resulted in $q$ decreasing with redshift, but only at $\sim2\sigma$ significance.}
}
\label{fig:qz}
\end{figure*}

\subsubsection{Supernovae}

We repeated the analysis for SNe, i.e., calculated the required dust yields per one massive star ($>8\,M_\odot$).
The yields are $\sim0.15$--$0.65\,M_\odot$ of dust per SN (the third row of Table~\ref{tab:dustyields}), consistent with the theoretical works (though without dust grain destruction implemented) of \citet{todini01} and \citet{nozawa03}; with a value predicted by \citet{dwek07} to account for dust in a $z\sim6.4$ quasar;  and with {\submm} observation of SN remnants \object{Cassiopeia A} \citep{dunne03,dunne09casA} and Kepler \citep{morgan03b,gomez09}. There is a debate about the latter results on \object{Cassiopeia A} and Kepler \citep[e.g.,][]{dwek04,krause04,gomez05,wilson05, blair07,sibthorpe09}, but if { the dust yields of \citet{dunne09casA} and \citet{gomez09} were correct and typical, then SNe would be efficient enough to account for the dust in { all} $z>4$ SMGs.}

However our,  required estimates are at least an order of magnitude higher than any other observed SN dust yields, which are typically in the range $\sim10^{-3}$--$10^{-2}\,M_\odot$
 \citep{green04,borkowski06,sugerman06,ercolano07,meikle07,rho08,rho09,kotak09,sakon09,sandstrom09,wesson09} and theoretically predicted dust masses able to survive in SN remnants \citep{bianchi07}.
This apparent difficulty in explaining dust production in $z>4$ SMGs can be resolved with a combination of two plausible effects.

Approximately half of the discrepancy can be accounted for with a top-heavy IMF giving more SNe per unit stellar mass \citep[both top-heavy and Salpeter IMFs have been claimed to reproduce the number counts of SMGs;][]{baugh05,fontanot07}. Changing the IMF slope from $\alpha=2.35$ to $\alpha=1.5$, consistent with values for low-mass star clusters \citep{scalo98} and a limit derived for a proto-star cluster \citep{sternberg98}, resulted in the required dust yield decreasing to $0.05$--$0.23\,M_\odot$ (the fourth row of Table~\ref{tab:dustyields}). 

The second possibility is that SNe provided only the dust seeds and that the bulk of the dust mass was accumulated during  grain growth in the  ISM \citep[e.g.,][]{draine03}.
The timescale of this process is typically less than a few$\mbox{}\times10$ Myr \citep{hirashita00,zhukovska08,draine09}, i.e., short enough to contribute significantly to the growth of dust mass in $z>4$ SMGs. { However, the extreme environments of $z>4$ SMGs may make it difficult for grains to grow before they are destroyed.} { The hypothesis of dust growth can be tested by investigation whether the grains formed by SN remnants are smaller than those present in $z>4$ SMGs.}

\subsection{The IR-Radio Correlation at $z>4$}
\label{sec:firr}

Despite the differences compared to the $1.7<z<3.6$ population (Section~\ref{sec:stars}), 
{
 the mean IR-to-radio luminosity ratio ($q\equiv\log[L_{\rm IR}/3.75\times10^{12}/I_{1.4}]$ with radio $K$-corrections assuming a slope of $-0.75$) for $z>4$ SMGs of $2.32\pm0.20$ 
is  consistent with the values derived for lower-redshift SMGs \citep{michalowski10smg}. At $z>4$ we find an offset of $\Delta q\sim-0.32$ (a factor of $\sim2.1$) from the local value of $q=2.64$ \citep{bell03}, though due to the small sample, this offset is significant only at $\sim1.6\sigma$ level. 

Hence,  with our multi-wavelength approach we confirm the results of 
\citet{murphy09b}, who derived the mean $q=2.16\pm0.28$  for $z>4$ SMGs. Moreover, our estimate is consistent with a value of $2.41\pm0.20$ derived for $250\,\micron$-selected galaxies at $z<3$ by \citet{ivison09}.

The offset of $\Delta q\sim-0.32$ is consistent with a hypothesis of \citet{lacki09} and \citet{lacki09b} that SMGs are ``puffy starbursts'' (vertically and radially extended galaxies with vertical scale heights $\sim1$ kpc) experiencing weaker 
bremsstrahlung and ionization
 losses 
  resulting in stronger radio emission. We cannot however exclude an AGN contribution boosting their radio fluxes. Since the redshifts of four out of six $z>4$ SMGs have been measured independently of radio detections, the radio excess cannot be a result of a bias against radio-faint sources.
}

In Figure \ref{fig:qz}, we show the $q$-values as a function of redshift for both $z>4$ SMGs discussed in this paper (red circles) and lower-redshift SMGs from \citet{michalowski10smg}. 
{ The IR-radio correlation of SMGs does not show any significant evolution in the redshift range $1.4<z<5.0$.}
{
A linear fit to all the data results in a low-significance ($\sim2\sigma$) indication that $q$ decreases with redshift \citep[as also found by][from a stacking analysis of $24\,\micron$-selected galaxies]{ivison09} in the form $q=(-0.078\pm0.038) z + (2.50\pm0.09)$.
}
To date, there are only two $q$ determinations at higher redshifts than presented here, namely $z=6.2$ and $6.42$ quasars \citep[][$q=1.8$--$2.2$]{carilli04,beelen06}.

\subsection{Gas-to-Dust Ratio at $z>4$}

Using the molecular gas mass estimates (based on CO[4--3] line observations) from \citet{schinnerer08} and \citet{daddi09,daddi09b} we derive gas-to-dust ratios of $M_{H_2}/M_d=73$, $47$, $22$, and $98$ for  \object{MMJ100054.48+023435.9}, \object{GN20}, \object{GN20.2a}, and \object{GN10}, respectively. { We note that these values are affected by significant uncertainties of the conversion from CO line strength to H$_2$ mass.} The mean value of $60$ is consistent with $54^{+14}_{-11}$ estimated for $z\sim1$--$3.5$ SMGs by \citet{kovacs06} using the CO survey of \citet{greve05}. SMGs at $z>4$ have one of the lowest gas-to-dust ratios compared to galaxies { selected at other wavelengths}, e.g., the Milky Way  \citep[$\sim90-400$;][]{sodroski97}, other spirals  \citep[$\sim1000 \pm 500$;][]{devereux90,stevens05}, the nuclear regions of local luminous IR galaxies, ultraluminous IR galaxies \citep[$120\pm28$;][]{wilson08}, and of local, far-IR-selected galaxies \citep[$\sim50$;][]{seaquist04} (all these results are based on IR and CO data, and are therefore directly comparable to our estimates). This is not surprising, since SMGs are selected by {\submm} emission. Similarly, a low value of $M_{H_2}/M_d=30$ was found by \citet{bertoldi03b} for $z=6.42$ quasar, but \citet{cox02} reported a higher value of $M_{H_2}/M_d=150$ for $z\sim4$ quasar. The small number of SMGs and quasars with derived gas-to-dust ratios hampers a comparison of these two samples.

\section{Conclusions}
\label{sec:conclusion}

We have analyzed the SEDs of six spectroscopically confirmed $z>4$ SMGs. Our results provide constraints on dust production at these early epochs of the evolution of the universe. 
We find that AGB stars
are efficient and fast enough to form the dust residing in three of these galaxies. However, for { the three remaining }  SMGs only SNe are efficient and fast enough. The high required SN dust yields hint at a possibility that their stars may be distributed according to a top-heavy IMF and/or that dust grains are substantially grown in the ISM. Since the majority of the stars in these
galaxies were formed on very short timescales, these properties are very likely to be  similar to those of the first galaxies beyond redshift $6$, which had been building up their stellar populations only for several hundred Myr, which elapsed since the big bang.

We present evidence that the IR-to-radio luminosity ratios of $z>4$ SMGs are consistent with that of lower-redshift SMGs and are offset from the local relation by a factor of $\sim2.1$.

A comparison of the $z>4$ SMGs with the lower-redshift sample, in particular their high SFRs, dust temperatures and fraction of stars formed during the ongoing starburst as well as low stellar masses, reveals that we start to see SMGs at earlier stages of their evolution. 

The improved mapping speed and sensitivity of the new SCUBA2 camera will enable studies of the evolutionary sequence of SMGs using much bigger and more homogeneously selected samples.
Moreover, the study of dust production presented here will be pushed forward with a synergy of {\it Herschel}, SCUBA2, and ultimately ALMA. These facilities will provide a broad wavelength coverage at the IR, which will allow accurate determination of dust temperatures and, in turns,  its mass.

\acknowledgments

We thank Joanna Baradziej, Amy Barger,  Rob Ivison and Peter Laursen
for discussion and comments; our referee for help with improving this paper; Kirsten Knudsen for kindly providing the data for SMMJ163555.5+661300 prior to publication; and
 Jorge Iglesias-P\'{a}ramo for providing his SED templates. 

M.J.M.~acknowledges support from The Faculty of Science, University of Copenhagen. The Dark Cosmology Centre is funded by the Danish National Research Foundation. 

\appendix
\section{Dust yield calculation}

We calculated the dust yield per star required to explain dust mass in a galaxy in { the} following way. 
In an IMF with $M_{\rm min}=0.15$, $M_{\rm max}=120\,M_\odot$, and a slope $\alpha=2.35$ \citep[][or $\alpha=1.5$ for top-heavy IMF]{salpeter}, the number of stars with masses between $M_0$ and $M_1$ in the stellar population with a total mass of $M_*$ can be expressed as
\begin{equation}
N(M_0<M<M_1)=M_*\frac{\displaystyle\int_{M_0}^{M_1} M^{-\alpha} dM } {\displaystyle\int_{M_{\rm min}}^{M_{\rm max}} M^{-\alpha}  M dM},
\label{eq:nstars}
\end{equation}
where the denominator provides a normalization so that a total mass is equal to $M_*$. For SNe we assumed $M_0=8\,M_\odot$ and $M_1=M_{\rm max}=120\,M_\odot$, whereas for AGB stars: $M_0=2.5\,M_\odot$ and $M_1=8\,M_\odot$.

The average dust yield per star is equal to the dust mass divided by the number of stars, $N(M_0<M<M_1)$.


\input{ms.bbl}

\end{document}

%% file: phottab.tex
\begin{table}
\caption{Photometry Detections of  $z>4$ SMGs \label{tab:phot}	}
\begin{tabular}{lrrrrl}
\hline\hline
 & & \colhead{$\lambda_{\rm obs}$} & \colhead{Flux} & \colhead{Error} & \\
		SMG & \colhead{$z$} & \colhead{($\mu$m)} & \colhead{($\mu$Jy)} & \colhead{($\mu$Jy)} & Ref.\\
\hline
LESSJ033229.4-275619 & 4.760 & 0.597 & 0.072 & 0.006 & Coppin09 \\
LESSJ033229.4-275619 & 4.760 & 0.652 & 0.215 & 0.020 & Coppin09 \\
LESSJ033229.4-275619 & 4.760 & 0.771 & 0.350 & 0.013 & Coppin09 \\
LESSJ033229.4-275619 & 4.760 & 0.905 & 0.405 & 0.015 & Coppin09 \\
LESSJ033229.4-275619 & 4.760 & 3.600 & 2.860 & 0.060 & Coppin09 \\
\hline
\end{tabular}
\tablecomments{The measured fluxes (Column 4) are given with the corresponding $1\sigma$ errors (Column 5). The data for \protect\object{SMMJ163555.5+661300} have been corrected for lensing magnification of a factor of $5.5$. (This table is available in its entirety in a machine-readable form in the online journal. A portion is shown here for guidance regarding its form and content.)
}
\tablerefs{\citet{capak04,capak08,giavalisco04goods,wang04,wang07,wang09,frayer06,iono06,pope06,dannerbauer08,greve08,perera08,schinnerer08,casey09,chapin09,coppin09,daddi09,daddi09b,devlin09,knudsen09}}
\end{table}

%% file: limtab.tex
\begin{table}
\caption{Photometry Upper Limits of  $z>4$ SMGs \label{tab:lim}	}
\begin{tabular}{lrrrrl}
\hline\hline
 & & \colhead{$\lambda_{\rm obs}$} & \colhead{Flux} & \colhead{Error} & \\
		SMG & \colhead{$z$} & \colhead{($\mu$m)} & \colhead{($\mu$Jy)} & \colhead{($\mu$Jy)} & Ref.\\
\hline
LESSJ033229.4-275619 & 4.760 & 0.352 & 0.087 & 0.000 & Coppin09 \\
LESSJ033229.4-275619 & 4.760 & 0.433 & 0.013 & 0.000 & Coppin09 \\
LESSJ033229.4-275619 & 4.760 & 0.461 & 0.039 & 0.000 & Coppin09 \\
LESSJ033229.4-275619 & 4.760 & 0.538 & 0.059 & 0.000 & Coppin09 \\
SMMJ123711.7+622212 & 4.055 & 0.365 & 0.063 & 0.026 & Capak04\\
\hline
\end{tabular}
\tablecomments{When the error is equal to zero, the flux column denotes the $3\sigma$ upper limit. Otherwise it denotes formal flux at the position of an SMG. The data for \protect\object{SMMJ163555.5+661300} have been corrected for lensing magnification of a factor of $5.5$. (This table is available in its entirety in a machine-readable form in the online journal. A portion is shown here for guidance regarding its form and content.)}
\tablerefs{\citet{capak04,capak08,giavalisco04goods,wang04,wang07,wang09,frayer06,iono06,pope06,dannerbauer08,greve08,perera08,schinnerer08,casey09,chapin09,coppin09,daddi09,daddi09b,devlin09,knudsen09}}
\end{table}

%% file: grasilrestab.tex
\begin{sidewaystable}
\caption{Properties of the $z>4$ SMGs Derived from the SED Modeling \label{tab:grasilres}}
\begin{tabular}{lc rrrr r cccc crc c l}
\hline\hline
 &  & \multicolumn{4}{c}{SFR ($M_\odot$ yr$^{-1}$)} & \colhead{SSFR} & \colhead{$M_*$} & \colhead{$M_{\rm burst}/M_*$} & $M_*/L_K$ & $ M_{d}$ & $ L_{\rm IR}$ & \colhead{$T_d$} & \colhead{$A_V$} &  &     \\
\cline{3-6}
SMG & $z$ & \colhead{SED} & \colhead{UV} & \colhead{IR} & \colhead{radio} & \colhead{(Gyr$^{-1}$)}  & \colhead{($10^{11}M_\odot$)} & (\%) & ($M_\odot/L_\odot$) & ($10^{9}M_\odot$) & ($10^{13}L_\odot$) & \colhead{(K)} & \colhead{(mag)} & $q$ & AGN?	\\
(1)  & (2)              &  \colhead{(3)} & \colhead{(4)} &  \colhead{(5)} & \colhead{(6)} & \colhead{(7)} & \colhead{(8)} & (9) & (10) & \colhead{(11)} & \colhead{(12)} & (13) & (14) & (15) & (16)\\
\hline
LESSJ033229.4-275619 & 4.760 &         1718 &           26 &         1150 & 
        1636 & 1.82 & 6.3 & 12.5 & 0.68 & 0.7 & 0.7 & 38.7 & 2.01 & 2.36 & 
SB,spec\\
MMJ100054.48+023435.9 & 4.547 &         3316 &           60 &         2725 & 
        3548 & 14.38 & 1.9 & 82.3 & 0.30 & 0.4 & 1.6 & 71.5 & 1.77 & 2.40 & SB\\
SMMJ123711.7+622212(GN20,AzGN01) & 4.055 &         6120 &           33 & 
        4577 &         4498 & 5.61 & 8.2 & 33.5 & 0.64 & 1.1 & 2.7 & 61.3 & 2.55
 & 2.52 & \dots\\
SMMJ123709.5+622206(GN20.2a,AzGN1.2) & 4.051 &          891 &           27 & 
         841 &        10700 & 3.44 & 2.4 & 14.3 & 0.48 & 1.4 & 0.5 & 38.7 & 1.76
 & 1.41 & rad\\
SMMJ123633.8+621408(GN10,GN850-5,AzGN03) & 4.042 &         2877 &            0
 &         5031 &         1951 & 28.57 & 1.8 & 100.0 & 14.03 & 0.3 & 2.9 & 97.2
 & 7.84 & 2.93 & SB\\
SMMJ163555.5+661300 & 4.044 &          339 &            1 &          289 & 
         $<$483 & 3.40 & 0.9 & 18.1 & 0.95 & 0.2 & 0.2 & 45.1 & 2.91 & $>$2.29 & SB\\
\hline
\end{tabular}
\tablecomments{Column~(1):~SMG name \citep[alternative names from][]{wang04,wang09,pope06,daddi09,daddi09b,perera08,chapin09}. Column~(2):~redshift \citep{coppin09,capak08,daddi09,daddi09b,knudsen09}. Column~(3):~total star formation rate (SFR) for $0.15-120\,M_\odot$ stars averaged over the last 50 Myr derived from the SED model. Column~(4):~SFR from UV emission interpolated from the SED template \citep[using][]{kennicutt}. Column~(5):~SFR from IR emission (Column 12) used in all analysis throughout the paper \citep[using][]{kennicutt}. Column~(6):~SFR from radio emission derived directly from the radio data \citep[using][]{bell03}. Column~(7):~specific SFR$\mbox{}\equiv\mbox{SFR}_{\rm IR}/M_*$. Column~(8):~stellar mass. Column~(9):~Ratio of the mass of gas converted to star during the recent starburst episode to the total stellar mass.  Column~(10):~stellar mass to light ratio (luminosity at rest-frame K was interpolated using the best SED model). Column~(11):~dust mass. Column~(12):~total $8-1000$ \micron\, infrared luminosity. Column~(13):~dust temperature (we assumed emissivity index $\beta=1.3$). Column~(14):~Average extinction $A_V=2.5\log$($V$-band starlight unextinguished / $V$-band starlight observed). Column~(15):~IR-radio correlation parameter (Section~\ref{sec:firr}). Column~(16):~AGN flag ---  SB: X-ray identified starburst \citep[no X-ray detection;][]{coppin09,capak08,dannerbauer08,wang09,knudsen09}, as opposed to X-ray identified AGN \citep{alexander05}; spec: spectroscopically identified AGN \citep{coppin09}; rad: radio datapoint is more than $3\sigma$ above the starburst model.   }
\end{sidewaystable}

%% file: dustyieldstab.tex
\begin{table*}
\caption{Dust Yields Per Star Required to Explain Dust in $z>4$ SMGs \label{tab:dustyields}}
\centering
\begin{tabular}{lll cccccc}
\hline\hline
& & & \multicolumn{6}{c}{Dust Yields ($M_\odot$ Per Star)}\\
\cline{4-9}
\colhead{Dust Producer} & \colhead{IMF} & \colhead{Total Mass} & LESS & MM & GN20 & GN20.2a & GN10 & SMM  \\
\hline
AGB ($2.5<M<8M_\odot$) & Salpeter & $M_*$
 & 0.03
 & 0.06
 & 0.04
 & 0.17
 & 0.05
 & 0.07
\\
AGB ($2.5<M<8M_\odot$) & Salpeter & $M_*-M_{\rm burst}$
 & 0.04
 & 0.31
 & 0.06
 & 0.20
 & $\infty$
 & 0.09
\\
SN ($M>8M_\odot$) & Salpeter & $M_*$
 & 0.14
 & 0.22
 & 0.15
 & 0.65
 & 0.18
 & 0.27
\\
SN ($M>8M_\odot$) & Top-heavy & $M_*$
 & 0.05
 & 0.08
 & 0.05
 & 0.23
 & 0.06
 & 0.10
\\
\hline
\end{tabular}
\tablecomments{The IMF is either \citet{salpeter} with $\alpha=2.35$ or top-heavy with $\alpha=1.5$. The total mass indicates if the entire stellar mass ($M_*$) was used to calculate the number of stars (Equation~\ref{eq:nstars}), or if stars created during the ongoing starburst were excluded ($M_*-M_{\rm burst}$). The last six columns contain dust yields for all SMGs in the order given in Table~\ref{tab:grasilres}. Only the first parts of their names are given for brevity.  }
\end{table*}